\documentclass[doublecol]{epl2}
\usepackage{amssymb}
\title{Doping evolution of superconducting gaps and electronic densities of states in Ba(Fe$_{1-x}$Co$_{x}$)$_{2}$As$_{2}$ iron pnictides}
\shorttitle{Doping evolution of the superconducting gaps and electronic
densities of states in Ba(Fe$_{1-x}$Co$_{x}$)$_{2}$As$_{2}$} %Insert

\author{F. Hardy\inst{1} \and P. Burger\inst{1} \and T. Wolf\inst{1} \and R. A. Fisher\inst{2}
\and P. Schweiss\inst{1} \and P. Adelmann\inst{1} \and R. Heid\inst{1} \and R. Fromknecht\inst{1} \and R. Eder\inst{1} \and D. Ernst\inst{1} \and H. v. L\"ohneysen\inst{1,3} \and C. Meingast\inst{1}}
\shortauthor{F. Hardy \etal}

\institute{
\inst{1} Karlsruher Institut f\"ur Technologie, Institut f\"ur Festk\"orperphysik, 76021 Karlsruhe, Germany\\
\inst{2} Lawrence Berkeley National Laboratory, Berkeley CA 94720, USA\\
\inst{3} Karlsruher Institut f\"ur Technologie, Physikalisches Institut, 76128 Karlsruhe, Germany\\
}

\abstract{An extensive calorimetric study of the normal- and superconducting-state properties of Ba(Fe$_{1-x}$Co$_{x}$)$_{2}$As$_{2}$ is presented for 0 $<$ x $<$ 0.2. The normal-state Sommerfeld coefficient increases (decreases) with Co doping for x $<$ 0.06 (x $>$ 0.06), which illustrates the strong competition between magnetism and superconductivity to monopolize the Fermi surface in the underdoped region and the filling of the hole bands for overdoped Ba(Fe$_{1-x}$Co$_{x}$)$_{2}$As$_{2}$. All superconducting samples exhibit a residual electronic density of states of unknown origin in the zero-temperature limit, which is minimal at optimal doping but increases to the normal-state value in the strongly under- and over-doped regions. The remaining specific heat in the superconducting state is well described using a two-band model with isotropic {\it s}-wave superconducting gaps.}

\begin{document}
\maketitle
\section{Introduction}
Despite the fact that the theoretical background has been available since the late 50's with the pioneering papers of {Suhl \etal}~\cite{Suhl59} and Moskalenko~\cite{Moskalenko59}, multiband superconductivity (MBSC) emerged as an unanimously accepted phenomenon only after the discovery of the MgB$_{2}$ superconductor in 2001~\cite{Nagamatsu01}. Rapidly, calorimetric signatures, like the excess specific heat observed at low temperature (with respect to the single-band BCS curve)~\cite{Bouquet01PRL}, the initial rapid rise of the mixed-state heat capacity with magnetic field~\cite{Bouquet02PRL}, and the anomalous positive curvature of the upper critical field~\cite{Lyard02} provided the most convincing evidence of its existence and are now textbook hallmarks of the existence of two gaps. Later, significant interband contributions to the Eliashberg function, reminiscent of MBSC, were experimentally detected using tunneling experiments~\cite{Geerk05}. Since then, the occurrence of MBSC has been discussed for many different compounds including heavy fermions~\cite{Seyfarth08}, cobaltates~\cite{Oeschler08}, chalcogenides~\cite{Huang07}, A15 compounds~\cite{LortzA15}, and the recently discovered iron-pnictide family. The aforementioned characteristic signatures of MBSC are less pronounced in iron pnictide superconductors since the interband coupling, promoted by the strong Fermi-surface nesting, is more important than in MgB$_{2}$ leading to a smaller gap anisotropy. Although, the existence of several energy gaps appears to be accepted, the precise symmetry of the superconducting order parameter and its evolution with doping remain controversial. Specific-heat~\cite{HardyPRB09,Popovich10} and ARPES~\cite{Terashima09,Ding08} studies agree on isotropic {\it s}-wave gaps in Ba(Fe$_{1-x}$Co$_{x}$)$_{2}$As$_{2}$ and Ba$_{1-x}$K$_{x}$Fe$_{2}$As$_{2}$, but differ on the gap amplitudes. On the other hand, Raman scattering~\cite{Muschler09} and heat transport measurements~\cite{Tanatar10}, which are both interpreted in terms of accidental nodes (and/or gap minima), show contradictory behavior with doping.

In this Letter, we present an extensive thermodynamic study of the evolution of the normal- and superconducting state properties of Ba(Fe$_{1-x}$Co$_{x}$)$_{2}$As$_{2}$ single crystals with Co concentration in the range 0 $<$ x $<$ 0.2. We find that the normal-state Sommerfeld coefficient ($\gamma_{n}$) first increases with Co doping up to x = 0.06 (where T$_{c}$ is maximal), which is related to the gradual suppression of the spin-density-wave (SDW) state; $\gamma_{n}$ then decreases for x $>$ 0.06, which we attribute to a gradual filling of the hole bands with increasing Co content. This illustrates the strong competition between magnetism and superconductivity to monopolize the Fermi surface (FS) in Ba(Fe$_{1-x}$Co$_{x}$)$_{2}$As$_{2}$. Further, we find that all superconducting samples exhibit a residual electronic density of states ($\gamma_{r}$) in the zero-temperature limit, which is minimal at optimal doping but increases up to the normal-state value in the strongly under- and overdoped states. We show that this residual term is intrinsic and not due to, {\it e.g.}, residual flux or oxidation products in the crystals. The observed values of $\gamma_{r}$ imply that the superconducting fraction strongly depends on doping and is only close to unity near optimal doping. We compare this behavior with that of the cuprates, which show a similar tendency. The superconducting-state specific heat is well described using a two-band model with isotropic {\it s}-wave gaps. The smaller gap ratio $\Delta_{1}$(0)/k$_{B}$T$_{c}$ is nearly independent of doping, whereas the large gap ratio $\Delta_{2}$(0)/k$_{B}$T$_{c}$ shows a tendency to strong coupling at optimal doping.
\begin{figure}[h]
\begin{center}
\includegraphics[width=8.5cm]{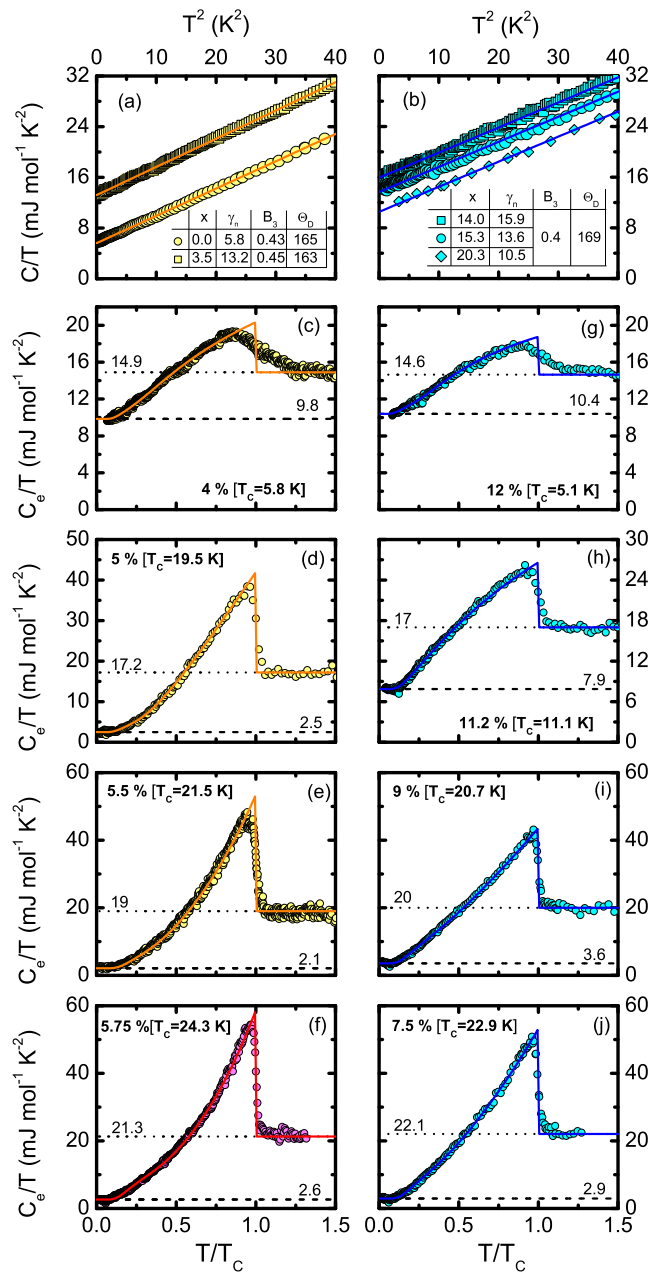}% Here is how to import EPS art
\caption{\label{fig:Fig1} Specific heat of underdoped (yellow symbols), optimally doped (magenta symbols) and overdoped (blue symbols) Ba(Fe$_{1-x}$Co$_{x}$)$_{2}$As$_{2}$ single crystals. (a)-(b) C/T of non-superconducting samples as a function of T$^{2}$. Lines are fit to the Debye model with $\gamma_{n}$, B$_{3}$, and $\Theta_{D}$ given in mJ mol$^{-1}$ K$^{-2}$, mJ mol$^{-1}$ K$^{-2}$, and K, respectively. (c)-(j) Electron specific heat C$_{e}$/T of several superconducting samples. Dotted and dashed lines represent $\gamma_{n}$ and $\gamma_{r}$, respectively (numerical values are given). Solid lines are fits to the two-band $\alpha$-model.}
\end{center}
\end{figure}
\section{Experimental details}
Single crystals, with typical mass of 5-10 mg, were grown from self-flux using prereacted FeAs and CoAs powders mixed with Ba. Details of the crystal synthesis are described elsewhere~\cite{HardyPRL09,HardyPRB09}. The precise composition of all samples was determined by energy dispersive x-ray spectroscopy and complemented by 4-circle
diffractometry. As reported by other groups, our characterization shows that the effective Co amount in the sample represents only 75 \% of the nominal flux Co content. The specific heat was measured by relaxation calorimetry down to 0.4 K, with the $^{3}$He-PPMS from Quantum Design. We have observed a slow degradation of the samples with time (over several months), which resulted in the progressive growth of a low-temperature Schottky anomaly. Therefore, all the measurements presented in this article were performed on crystals which were cooled down and measured within less than 10 hours after opening the SiO$_{2}$ ampoule.

\section{Specific-heat measurements}
Figures \ref{fig:Fig1}(a)-(b) display the specific heat plotted as C/T as a function of T$^{2}$ (above 0.4 K) of non-superconducting underdoped (x $<$ 0.04) and overdoped (x $\geq$ 0.14) single crystals below 6 K. In contrast to other reports, we do not observe any Schottky-like contribution, revealing the high quality of these samples. The heat capacity accurately follows the standard behavior, C/T=$\gamma_{n}$+B$_{3}$T$^{2}$, where $\gamma_{n}$ and B$_{3}$ are the normal-state Sommerfeld and Debye terms, respectively. The Debye temperature changes by less than 3 \% between the low-doping orthorhombic and the overdoped tetragonal phases confirming that the substitution of Fe by Co barely affects the phonon spectrum, in accord with neutron scattering measurements~\cite{Reznik09}. Figures \ref{fig:Fig1}(c)-(j) show the electronic specific-heat contribution (C$_{e}$) of superconducting crystals with 0.04 $\leq$ x $\leq$ 0.12, obtained by subtracting the lattice specific heat from the raw data following the procedure given in Ref.~\cite{HardyPRB09}. For overdoped (underdoped) single crystals, we used the lattice contribution derived from the x = 0.153 (x = 0.03) sample. All data comply with entropy conservation, {\it i.e.}, $\int_{0}^{T_{c}} \left( C_{e}/T \right) dT$ = $\gamma_{n}$$T_{c}$. Except for the x = 0.04 and x = 0.12 curves, the superconducting transitions at T$_{c}$ are remarkably sharp, indicating little inhomogeneity in the crystals. Nevertheless, our results show that C$_{e}$/T does not extrapolate to zero at T = 0, but exhibits a residual normal-state-like contribution, $\gamma_{r}$. In the following, we discuss in detail the doping evolution of (i) the normal-state electronic density of states (EDOS) and the resulting phase diagram, (ii) the superconducting gaps derived from fitting the specific heat with a two-band model (solid lines in Fig. \ref{fig:Fig1}(c)-(j)) and (iii) the residual electronic density of states (RDOS).

\section{(x,T) phase diagram and normal-state EDOS}
Fig. \ref{fig:Fig2}(a) shows the phase diagram derived from our specific-heat data together with thermal-expansion measurements performed on the same samples (Ref.~\cite{Meingast10}). Our phase diagram is very similar to previous reports, except that superconductivity disappears at a lower Co concentration (x $\approx$ 0.125) than previously reported from transport and magnetization measurements~\cite{Chu09,Rullier09,Katayama09,Canfield09}. We also observe that superconductivity emerges already in the magnetic phase, at a concentration x $\approx$ 0.04, and T$_{c}$ increases to 24.2 K at x $\approx$ 0.575. This increase is accompanied by a rise of $\gamma_{n}$ from 5.6 mJ mol$^{-1}$ K$^{-2}$ at x = 0 to about 22 mJ mol$^{-1}$ K$^{-2}$ close to the optimal concentration (see Fig. \ref{fig:Fig2}(b)), which we attribute to the closing of the SDW gap with Co doping. Our data show that the decrease of $\gamma_{n}$ with increasing x is linear for x $>$ 0.0575 and extrapolates for x = 0 to a value of $\approx$ 27 mJ mol$^{-1}$ K$^{-2}$ (dashed line in Fig. \ref{fig:Fig2}(b)), which we interpret as the high-temperature estimate of the Sommerfeld coefficient for T $>$ T$_{N}$, {\it i.e.}, in the paramagnetic phase of BaFe$_{2}$As$_{2}$. This indicates that the EDOS is reduced by a factor of six due to the FS reconstruction caused by the SDW. An increase of $\gamma_{n}$ implies that a larger fraction of the Fermi surface becomes available for superconductivity, which naturally explains the increase of T$_{c}$ and $\Delta$C/$\gamma_{n}$T$_{c}$ with Co doping in the underdoped region. In fact, it was recently shown that SDW and superconducting states coexist and compete for the same FS in this region of the phase diagram~\cite{Fernandes10,Fernandes10-2}. On the other hand, for x $>$ 0.575, both $\gamma_{n}$ and T$_{c}$ decrease as the system moves away from the magnetic instability. In Al- and C- doped MgB$_{2}$~\cite{Delapena09}, the reduction of $\gamma_{n}$ with electron doping is well understood by the filling of the 2D $\sigma$ hole bands and this interpretation may also hold for overdoped Ba(Fe$_{1-x}$Co$_{x}$)$_{2}$As$_{2}$. Indeed, ARPES measurements~\cite{Sekiba09} have shown that x = 0.10 superconducting samples have hole bands only partially filled, whereas the same bands are completely shifted below the Fermi level E$_{F}$ when the Co content is increased to x = 0.15 where T$_{c}$ is zero. Therefore, the collapse of superconductivity around x = 0.13 is likely related to the disappearance of the interband nesting condition that connect hole and electron bands. Moreover, in doped MgB$_{2}$, the reduction of the EDOS at E$_{F}$ also leads to a weakening of the electron-phonon coupling resulting in the disappearance of superconductivity~\cite{Delapena09}. 

\section{Two-gap analysis}
Although, in principle, three or more bands should be used to describe the thermodynamics of the Fe pnictides, simple two-band models have been successfully applied to describe a variety of experimental data. In the following, we therefore use a two-gap model to analyze the heat-capacity data, which has the advantage of a minimal number of fit parameters. As shown in Figs. \ref{fig:Fig1}(c)-(j), C$_{e}$ is very well represented, for all superconducting samples, by two-gap fits similar to those made for MgB$_{2}$~\cite{Bouquet01EPL}.
\begin{figure}[t]
\begin{center}
\includegraphics[width=8.5cm]{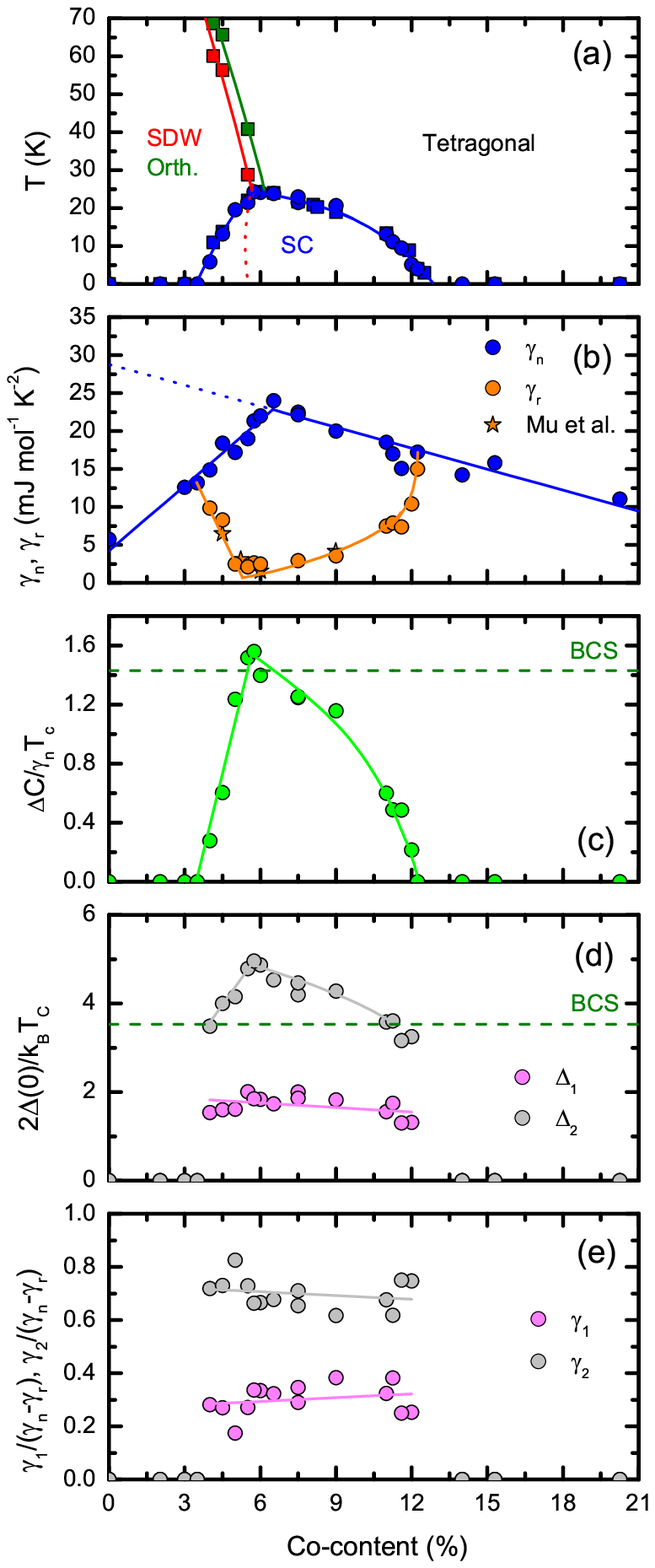}% Here is how to import EPS art
\caption{\label{fig:Fig2} (a) Phase diagram of Ba(Fe$_{1-x}$Co$_{x}$)$_{2}$As$_{2}$ derived from our specific-heat (circles) and thermal-expansion (squares) measurements. (b) Sommerfeld coefficient $\gamma_{n}$ and the residual density of states $\gamma_{r}$. Measurements of Mu {\etal}~\cite{Mu10} are shown for comparison. (c) Normalized specific-heat jump. The dashed line represents the BCS value $\Delta $C/$\gamma_{n}$T$_{c}$=1.43. (d)-(e) Reduced gaps and band densities of states (normalized to 1 mole of superconducting material) derived from the two-band analysis. Solid lines are guides to the eye.}
\end{center}
\end{figure}
In these fits, C$_{e}$ is the sum of three contributions: the residual term, $\gamma_{r}$T, and the contributions of the two gapped components of the superconducting condensate derived from the two bands, {\it i} = 1, 2. The temperature dependences of the superconducting-state contributions are those given by the $\alpha$-model~\cite{Padamsee73}, in which the weak-coupling BCS temperature dependence of the gaps is assumed, but the T = 0 value, $\Delta_{i}$(0), is an adjustable parameter represented by $\alpha_{i}$ $\equiv$ $\Delta_{i}(0)/$k$_{B}$T$_{c}$ (in the BCS weak-coupling limit $\alpha$(0) = 1.764). The amplitudes of these contributions are proportional to $\gamma_{i}$, with the following constraint: $\gamma_{1}$ + $\gamma_{2}$ = $\gamma_{n}$ - $\gamma_{r}$. Thus, the fits introduce three adjustable parameters, $\alpha_{1}$, $\alpha_{2}$, and $\gamma_{1}$/$\gamma_{2}$. 

Fig. \ref{fig:Fig2}(c) shows the specific-heat ratio, $\Delta $C/$\gamma_{n}$T$_{c}$, as a function of Co content. At the optimal concentration, it reaches a value of about 1.6, which slightly exceeds the weak-coupling BCS value, but is significantly smaller than 2.5 reported for Ba$_{0.68}$K$_{0.32}$Fe$_{2}$As$_{2}$~\cite{Popovich10}. The evolution of $\Delta $C/$\gamma_{n}$T$_{c}$ and $\gamma_{r}$ with doping is strongly correlated, {\it i.e.} the largest (smallest) $\Delta$C/$\gamma_{n}$T$_{c}$ occurs for the smallest (largest) $\gamma_{r}$. This indicates that the strong doping dependence of $\Delta$C/$\gamma_{n}$T$_{c}$ is due primarily to the presence of $\gamma_{r}$. Indeed, after normalizing $\Delta $C to the superconducting fraction ($\gamma_{n}$-$\gamma_{r}$) instead of $\gamma_{n}$, the ratio $\Delta$C/($\gamma_{n}-\gamma_{r}$)T$_{c}$ exhibits a weaker doping dependence than $\Delta$C/$\gamma_{n}$T$_{c}$, which likely reflects the evolution of $\Delta_{2}$(0)/k$_{B}$T$_{c}$ with x. We note that recent specific-heat measurements by Bud'ko {\etal}~\cite{Budko09} have reported that $\Delta$C/T$_{c}$ $\propto$ T$_{c}^{2}$ for Ba(Fe$_{1-x}$Co$_{x}$)$_{2}$As$_{2}$ and Ba(Fe$_{1-x}$Ni$_{x}$)$_{2}$As$_{2}$, which was subsequently interpreted by Zaanen~\cite{Zaanen09} in terms of a quantum critical metal undergoing a pairing instability. We find that the same scaling relation holds, but with an exponent closer to $\approx$ 1.5 (not shown), and it is not clear whether this theoretical approach can explain the present data. In particular, any valid theory needs to explain that $\Delta$C is reduced because of the large $\gamma_{r}$, and also needs to consider that in a two-band superconductor, the size of the jump is governed by both the gap anisotropy, $\Delta_{2}$/$\Delta_{1}$, and the ratio of the different band densities of states, $\gamma_{2}$/$\gamma_{1}$. 

Figures \ref{fig:Fig2}(d)-(e) summarize the fit parameters derived from the isotropic two-band analysis. Except for the extremely underdoped and overdoped samples, the gap values (see Fig. \ref{fig:Fig2}(d)) conform to the theoretical constraints that, for a weakly coupled two-band superconductor~\cite{Kresin90}, one gap must be larger than the BCS value and the other smaller. The largest gap values are obtained at optimal doping with a small gap 2$\Delta_{1}(0)$/k$_{B}$T$_{c}$ $\approx$ 1.9 and a major one with 2$\Delta_{2}(0)$/k$_{B}$T$_{c}$ $\approx$ 5. Recently, the same analysis was successfully applied to the heat capacity of the close to optimally doped Ba$_{0.68}$K$_{0.32}$Fe$_{2}$As$_{2}$ by Popovich{\etal}~\cite{Popovich10}. They found a comparable 2$\Delta_{1}(0)$/k$_{B}$T$_{c}$ $\approx$ 2.2, whereas the larger ratio, 2$\Delta_{2}(0)$/k$_{B}$T$_{c}$ $\approx$ 6.6, is substantially higher, which was interpreted as a possible sign of strong coupling.

Fig.\ref{fig:Fig2}(e) illustrates that $\gamma_{2}$ is always greater than $\gamma_{1}$ indicating that the major gap develops around the Fermi surface sheet that exhibits the largest EDOS, while in a two-band model it is theoretically expected that $\gamma_{2}$/$\gamma_{1}$$\propto$$\sqrt{\Delta_{1}/\Delta_{2}}$ in the limit of pure interband coupling within the weak-coupling limit~\cite{Dolgov09}. These findings imply that either intraband couplings are not negligible or that more than two bands are involved in Co-doped 122 pnictides. As pointed out by Benfatto {\etal}~\cite{Benfatto09}, a minimal model of three bands is indeed required to describe the superconducting-state properties, in agreement with ARPES measurements in Ba$_{0.6}$K$_{0.4}$Fe$_{2}$As$_{2}$~\cite{Ding08} which have revealed the existence of three energy gaps. Our data can also be fitted with three gaps, but this would necessitate five fitting parameters. Our present analysis can be qualitatively reconciled with photoemission data by assuming that a small gap (2$\Delta_{1}(0)$/k$_{B}$T$_{c}$ $\approx$ 1.9) exists on the quasi-3D hole band ($\alpha$ band), while two large gaps of equal amplitude (2$\Delta_{2}(0)$/k$_{B}$T$_{c}$ $\approx$ 5) develop around the two strongly nested sheets, {\it i.e.}, the hole-like sheets ($\beta$ bands) and the electron-like bands ($\gamma$ bands). Recent first-principles calculations by Nakai {\etal}~\cite{Nakai10} support this assumption. In this context, $\gamma_{1}$ represents the density of states of the $\alpha$ band while $\gamma_{2}$ consists of contributions from the 2D $\beta$ and $\gamma$ sheets. In the following, we discuss our data using this scheme.

\begin{figure}[h]
\begin{center}
\includegraphics[width=8.5cm]{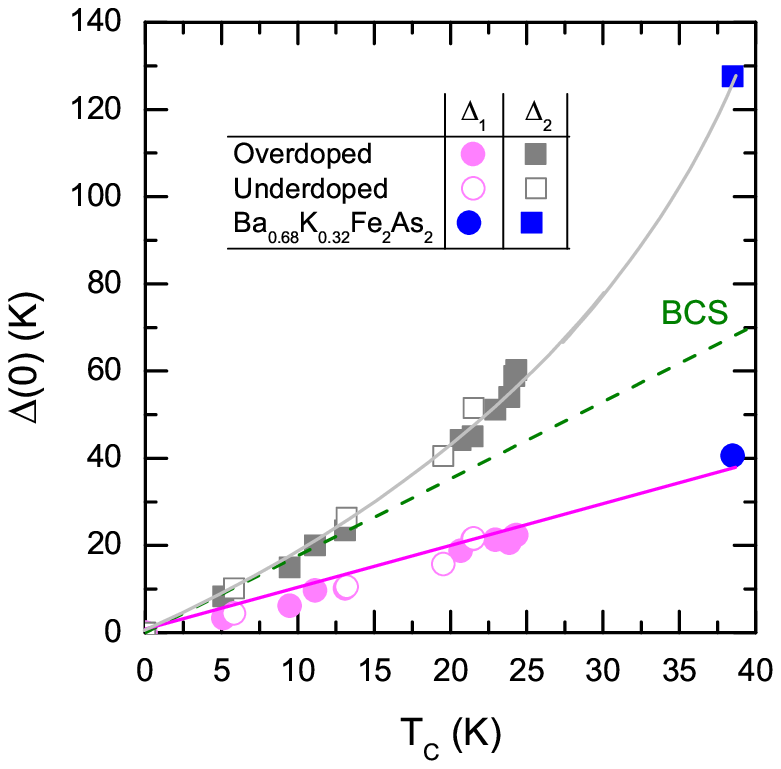}% Here is how to import EPS art 
\caption{\label{fig:Fig3} Evolution of the superconducting gaps with critical temperature T$_{c}$ for Ba(Fe$_{1-x}$Co$_{x}$)$_{2}$As$_{2}$. Filled and empty symbols represent overdoped and underdoped samples, respectively. Gap amplitudes of Ba$_{0.68}$K$_{0.32}$Fe$_{2}$As$_{2}$~\cite{Popovich10} are also shown (blue symbols). Lines are guide to the eye.} 
\end{center}
\end{figure}

In Fig.\ref{fig:Fig3} we plot both gap amplitudes as a function of T$_{c}$ together with values for Ba$_{0.68}$K$_{0.32}$Fe$_{2}$As$_{2}$~\cite{Popovich10}. We find that (i) the small gap increases linearly with T$_{c}$, (ii) the large gap increases stronger than linearly with T$_{c}$, and (iii) the same behavior occurs for both under- and overdoped samples. Additionally, the K-doped data fit nicely onto the same curves. In the limit of T$_{c}$$\rightarrow$ 0, the larger gap $\Delta_{2}(0)$ converges to the BCS value, $\Delta_{BCS}(0)$ = 1.764k$_{B}$T$_{c}$, which has previously been reported also for Mg$_{1-x}$Al$_{x}$B$_{2}$~\cite{Gonnelli07}. In the latter system, the coupling strength and the interband scattering between the 2D $\sigma$ bands (which exhibit the larger gap) and the 3D $\pi$ bands are weak because of their difference in character. By analogy, the same argument can also explain the difference in gap amplitude of the quasi-3D $\alpha$ band and the almost 2D $\beta$ and $\gamma$ bands in Ba(Fe$_{1-x}$Co$_{x}$)$_{2}$As$_{2}$.  Figure \ref{fig:Fig3} shows a tendency for strong-coupling effects for the large gap, {\it i.e.}, the compounds with the highest T$_{c}$, especially for the K-doped superconductor. Recent calculations~\cite{Neupane10} have shown that the Lindhard function, which provides a measure of the degree of nesting, is much higher for K- than for Co-doped Ba122, and this strong coupling may thus be related to the degree of nesting. Taken together, our data and analysis are fully consistent with superconductivity driven by interband coupling between the nested $\beta$ and $\gamma$ bands.

\section{Residual density of states, $\gamma_{r}$}
We now focus our discussion on the RDOS shown in Fig. \ref{fig:Fig2}(b). Sizeable $\gamma_{r}$ values appear to be a general feature of specific-heat measurements on Co-doped 122 iron pnictides~\cite{HardyPRB09,Mu10,Gofryk10}, and the first question to be addressed is: Is this an intrinsic or an extrinsic ({\it e.g.}, residual flux, a second phase, or generally poor sample quality) effect? The systematic doping dependence of $\gamma_{r}$ shown in Fig. \ref{fig:Fig2}(b) strongly suggests an intrinsic origin.  This is supported by recent, albeit sparse data of Mu {\etal}~\cite{Mu10} (see Fig. \ref{fig:Fig2}(b)), which are practically identical in value to ours.  Intrinsic RDOS is usually not observed in classical BCS superconductors, not even for dirty MgB$_{2}$~\cite{Putti03,Putti06,FisherAIP06}. On the other hand, many unconventional superconductors appear to exhibit such a term.  For example, even the best YBCO samples near optimal doping have $\gamma_{r}$/$\gamma_{n}$ of rougly 10 \% ~\cite{Wang01}, the origin of which remains unclear.  Remarkably, an anticorrelated behavior between $\gamma_{r}$ and $\gamma_{n}$ similar to our results has been reported by Koike {\it et al.}~\cite{Koike08} in La$_{2-x}$Sr$_{x}$CuO$_{4}$, which was argued to result from microscopic phase separation into SC and non-SC metallic regions. This could also explain our results for Ba(Fe$_{1-x}$Co$_{x}$)$_{2}$As$_{2} $.  A very similar explanation has been used in the context of penetration-depth measurements, where the term 'Swiss cheese' was used to describe the mixture of normal and superconducting regions~\cite{Uemura89}.

On the other hand, $\gamma_{r}$ could also be explained by disorder, which is likely to be particularly important in unconventional superconductors with a sign-reversing order parameter. Recent NMR measurements~\cite{Ning08} have shown that cobalt substitution does not induce any local moment, hence Co can be considered as a non-magnetic scatterer. Moreover, it was argued that the observed power-law behavior of the penetration depth and the NMR spin-lattice relaxation time can be reconciled with an {\it s}-wave order paramater only if the scattering by non-magnetic impurities is taken into account~\cite{Parker08,Vorontsov09}. As shown by Onari and Kontani~\cite{Onari09}, interband scattering by non-magnetic impurities causes only a moderate drop of T$_{c}$ for the {\it s$_{++}$} state, while for a sign-reversing order parameter, like the {\it s$_{+-}$} state, strong pair-breaking effects, including a drastic suppression of T$_{c}$ and the emergence of prominent in-gap states, are expected. Thus, if $\gamma_{r}$ is due to disorder, then the order parameter is likely to be {\it s$_{+-}$}. However, it is not clear how disorder alone (due to Co doping) can explain the minimum of $\gamma_{r}$ at optimal doping, since naively one would expect $\gamma_{r}$ to continually increase with x.  

In summary, from an extensive thermodynamic study of Ba(Fe$_{1-x}$Co$_{x}$)$_{2}$As$_{2}$ high-quality single crystals we find that the normal-state Sommerfeld coefficient first increases with Co doping up to x = 0.06 (where T$_{c}$ is maximal) and then decreases for x $>$ 0.06.  The maximum of $\gamma_{n}$ at x = 0.06 arises due to the gradual suppression of the spin-density wave state (SDW) for x $<$ 0.06 and to a continuous filling of the hole bands for x $>$ 0.06 with increasing Co content.  Our results, thus, illustrate the strong competition between magnetism and superconductivity to monopolize the Fermi surface in Ba(Fe$_{1-x}$Co$_{x}$)$_{2}$As$_{2}$. All our superconducting samples exhibit a residual electronic density of states in the zero-temperature limit, which is minimal at optimal doping but increases up to the normal-state value in the strongly under- and overdoped regions of the phase diagram. This residual term is likely to have an intrinsic origin and is not due to, {\it e.g.}, residual flux in the crystals. The values of $\gamma_{r}$ imply that the superconducting fraction is a strong function of doping and is only close to unity near optimal doping, which is similar to the behavior of cuprates and warrants further theoretical study.  A detailed analysis of the specific heat shows that the superconducting state is described well using a two-band model with isotropic {\it s}-wave superconducting gaps, which are interpreted in terms of a more microscopic three-band picture.  At optimal doping there is a tendency to strong coupling for the larger of the two gaps, whereas the smaller gap is nearly independent of doping. 

%\section{Section title}
%Insert here the text.
%See fig.~\ref{fig.1}, table~\ref{tab.1} and eq.~(\ref{eq.1}).
%See also~\cite{b.a,b.b}.
%\begin{equation}
%\label{eq.1}
%0\neq1
%\end{equation}

%\begin{figure}
%\onefigure{epl-template.eps}
%\caption{Figure caption.}
%\label{fig.1}
%\end{figure}

%\begin{table}
%\caption{Table caption.}
%\label{tab.1}
%\begin{center}
%\begin{tabular}{lcr}
%first & table & row\\
%second & table & row
%\end{tabular}
%\end{center}
%\end{table}

\acknowledgments
%Insert here the text.
This work is part of the DFG priority program (SPP 1458) "High-Temperature Superconductivity in Iron Pnictides" funded by the Deutsche Forschungsgemeinschaft (DFG).

\end{document}